\documentclass[twocolumn,prc, showpacs, superscriptaddress,nofootinbib,floatfix, amsfonts, aps]{revtex4}

\usepackage{amssymb}
\usepackage{setspace, placeins}
\usepackage{amsmath}
\usepackage{bm}
\usepackage{amsmath}
\usepackage{graphicx}
\usepackage{mathrsfs}
\usepackage{footmisc}
\usepackage{multirow}
\usepackage{graphicx}
\usepackage{booktabs, ctable}
\usepackage{xcolor, color, framed, ulem}

\begin{document}
\title{Hadronic Effects on Charmoium Elliptic Flows in Heavy-Ion Collisions}
\author{Baoyi Chen}
\author{Liu Jiang}
\author{Yunpeng Liu}
\email{yunpeng.liu@tju.edu.cn }
\affiliation{Department of Physics, Tianjin University, Tianjin 300350, China}

\date{\today}

\begin{abstract}
Transport and Langevin equations are employed to 
study hadronic medium effects on charmonium elliptic flows in heavy-ion 
collisions. In Pb-Pb collisions, 
the anisotropic energy density of the quark-gluon plasma (QGP)
in the transverse plane is transformed into hadron momentum anisotropy 
after the phase transition. 
Charmonia with high transverse momentum $p_T$ are produced via  
the primordial hard process and 
undergo different degrees of 
dissociation along different paths in the QGP. 
They then scatter with light hadrons in the hadron phase. 
Both contributions to the charmonium elliptic flows are studied at moderate and high 
transverse momenta. 
The elliptic flows of the 
prompt $J/\psi$ are found to be considerably enhanced at high transverse momentum 
when the charmonium diffusion coefficients in the hadronic medium 
are parametrized through the geometry scale approximation. 
This hadronic medium effect is 
negligible for quarkonia with larger mass such as bottomonia. 

\end{abstract}
\pacs{25.75.-q, 12.38.Mh, 14.40.Pq}
\maketitle

\section{introduction}
Charmonium consists of charm and anti-charm 
quarks. It is produced in parton hard scatterings with large momentum transfer. 
Charmonium 
has been considered as a clean probe of 
the hot dense medium created in relativistic 
heavy-ion collisions~\cite{Matsui:1986dk}.   
In the quark-gluon plasma (QGP), 
thermal light partons can dissociate charmonium states 
through inelastic collisions and the color screening effect, whereby the heavy quark 
potential in charmonium 
is screened by the light partons~\cite{Grandchamp:2001pf,Zhu:2004nw,Satz:2005hx, 
Blaizot:2015hya,Yao:2018sgn, Yao:2018nmy}. 
The charmonium dissociation rates induced by the two mechanisms are 
positively correlated to 
the QGP energy density~\cite{Chen:2019qzx}. At the Large Hadron Collider (LHC), 
QGP with large energy density dominates the abnormal suppression of charmonium yields 
compared with the effects of hadronic medium dissociations. 
Therefore, hadron phase contributions have been usually neglected in the analysis of the quarkonium 
nuclear modification factor $R_{AA}$ 
in the heavy-ion collisions at LHC collision energies. Furthermore, the hadronic medium 
produced after the violent expansion of QGP carries larger collective flows. 
This can affect the momentum distribution of charmonia through elastic scattering.

In nucleus-nucleus collisions, some charmonia are also produced through the 
combination of uncorrelated charm and anti-charm quarks in the 
QGP; this process is usually referred to as ``regeneration''~\cite{Thews:2000rj,Andronic:2003zv,
Yan:2006ve,Du:2015wha}. 
With numerous (anti-)charm quarks moving randomly in the QGP, their combination 
probability at the hadronization surface becomes non-negligible. From 
both theoretical and experimental studies, it has been found that more than half of the 
final $J/\psi$ in Pb-Pb central collisions 
are due to the combination of $c$ and $\bar c$
~\cite{Chen:2018kfo,Zhao:2011cv}. 
Being strongly coupled with the bulk medium, 
these charm quarks carry 
collective flows from the anisotropic expansion of the QGP~\cite{He:2014cla,Cao:2015hia}. 
This collective flow of 
charm quarks will be inherited by the regenerated charmonia~\cite{Acharya:2017tgv}. 
These regenerated charmonia are mainly located in the low $p_T$ region. 
In the high $p_T$ region where the regeneration contribution is negligible, charmonium yields 
are dominated by
the primordial production. 
Primordially produced charmonia undergo different magnitudes 
of dissociations when they move in different directions in the 
anisotropic QGP~\cite{Bhaduri:2018iwr}. 
This effect results in an anisotropic charmonium momentum distribution after they have moved out of the QGP. 
This contributes $\sim 2\%$ of the elliptic flow of $J/\psi$ in the 
high $p_T$ region in semi-central Pb-Pb collisions, which is considerably lower than that obtained from 
experimental data~\cite{Acharya:2017tgv}. 
In this work, the hadronic medium effects on charmonium elliptic flows are studied, particularly
in the high $p_T$ region. 
Charmonium 
trajectories are treated 
as Brownian motions in the hadronic medium, due to elastic scattering with thermal hadrons. 
This process can be described through the Langevin equation. The interactions between 
charmonium and the hadronic medium are parametrized via the diffusion coefficients in 
the Langevin equation~\cite{Mitra:2014ipa}.

The transport model is employed to study
charmonium evolutions in the QGP~\cite{Zhou:2014kka, Chen:2018kfo}, 
while the Langevin equation is used to model charmonium diffusions in the hadronic medium. 
This paper is organized as follows. The transport and Langevin models are 
introduced in Section II. 
The numerical results and analysis are presented in Section III. 
Section IV provides a summary. 

\section{transport model and langevin model}
The Boltzmann-type transport model has been developed to describe the phase space 
evolutions of 
heavy quarkonia in heavy-ion collisions. 
This model explains well and consistently most of the experimental data, including the 
spectrum and collective flows of charmonia and bottomonia~\cite{Zhu:2004nw, 
Zhou:2014kka, Chen:2016dke, Liu:2010ej, Yao:2020xzw}. The transport equation 
is written as: 
\begin{align}
\label{eq-trans}
&\partial_t f_\psi(t, {\bf x},{\bf p}) + {\bf v}_\psi\cdot 
{\bf \bigtriangledown_x} f_\psi(t, {\bf x}, {\bf p})
= - \alpha f_\psi(t, {\bf x}, {\bf p}) + \beta,
\end{align}
where $f_\psi(t,{\bf x},{\bf p})$ is the charmonium density in the phase space. 
The second term on the left-hand side represents the free streaming of charmonium 
with a constant velocity. In Eq.(\ref{eq-trans}), the elastic collision term in the QGP~\cite{Chen:2012gg} 
has been 
neglected. Hot medium effects on charmonia 
are parametrized through the $\alpha$ and $\beta$ functions. 
$\alpha$ is the charmonium dissociation rate in the QGP; it depends on 
the local temperatures and inelastic cross sections 
($g+\psi\rightarrow c+\bar c$) between 
the charmonia and thermal partons. 
This cross section is obtained from the  
operator production expansion (OPE)~\cite{Peskin:1979va,Bhanot:1979vb}. 
In the expression for the $\alpha$ dissociation rate, 
$J/\psi$ in-medium binding energy is taken to be 500 MeV. 
For the loosely bound excited states ($\chi_c, \psi^\prime$), 
the dissociation cross sections calculated using the OPE are no longer 
reliable. 
For this reason,
we choose in this work to obtain the dissociation rates of the excited
charmonium states by phenomenologically parameterizing them in terms
of a geometry scale with the ground state, as discussed below.
The dissociation rates of the charmonium excited 
states are obtained through the geometry scale with the ground state. 
These parameters regarding $\alpha$ are the same as those reported in a 
previous work~\cite{Chen:2018kfo}. The $\beta$ term represents the charmonium 
regeneration rate due to the process of $c$ and $\bar c$ combination. As this 
contribution is negligible for the $J/\psi$ production in the high $p_T$ region, 
it is not included in the following calculations of $J/\psi$ elliptic 
flows.  

The distributions of primordial charmonia in Pb-Pb collisions 
are obtained through a superposition of effective nucleon-nucleon collisions, according to:
\begin{align}
\label{eq-init}
f_{t=0}({\bf x}_T, {\bf p}_T, y|{\bf b})& =  T_A({\bf x}_T+{{\bf b}\over 2})
T_B({\bf x}_T-{{\bf b}\over 2})\nonumber \\
&\times  {d^2 \sigma_{J/\psi}^{pp}\over dy 2\pi p_T dp_T}
\mathcal{R}_{s}({\bf x}_T, {\bf p}_T, y),
\end{align}
where $T_{A(B)}$ is the thickness function of the nucleus A(B), 
$\bf b$ is the impact parameter, $y$ is the rapidity, 
and $d^2\sigma_{J/\psi}^{pp}/dy2\pi p_Tdp_T$ is the $J/\psi$ momentum distribution 
in pp collisions. $\mathcal{R}_s$ is the modification factor due to the shadowing effect; 
it is calculated via the EPS09 model~\cite{Eskola:2009uj}. 
At $\sqrt{s_{NN}}=5.02 $ TeV,  
the $J/\psi$ prompt differential cross section is 
taken to be $3.25\ \mu b$ in the 
forward rapidity~\cite{Abelev:2012kr,Abelev:2012gx}.

In the hadronic medium, all the 
collisions between charmonia and light hadrons are parametrized in terms of the 
drag coefficient and noise term in the Langevin equation.   
The classical Langevin equation~\cite{Chen:2017duy} for 
charmonium motion in a hadronic medium is written as:  
\begin{align}
{d{\bf  p}\over dt}=- \eta{\bf p}+{\bf \xi },
\label{fun-LG}
\end{align}
where $\bf p$ is the charmonium momentum, while 
$\eta$ and $\bf \xi$ are the drag force and the noise from the hadronic medium, respectively. 
For the simplicity of calculation, 
the classical form of the Langevin equation is assumed, and
the momentum dependence of both $\eta$ and $\bf \xi$ is neglected~\cite{Rapp:2009my}.  
${\bf \xi}$ satisfies the correlation relation:
\begin{align}
\langle \xi^{i}(t)\xi^{j}(t^\prime)\rangle =\kappa \delta ^{ij}\delta(t-t^\prime),
\end{align}
where $\kappa$ is the charmonium diffusion coefficient in the momentum space; it 
is connected with the 
spatial diffusion coefficient $\mathcal{D}_s^{\psi}$ 
through $\kappa=2T^2/\mathcal{D}_s^{\psi} $ 
in the classical limit. 
The drag force in the Langevin equation 
can then be determined via the Einstein  
fluctuation-dissipation relation $\eta\approx \kappa/2m_\psi T$. 
Here, $m_\psi$ is the charmonium mass, and $T$ is 
the local temperature of the hadronic medium. 
In this equation, only one parameter, namely $\mathcal{D}_s^{\psi}$ 
for charmonium, is as yet undetermined. 
The diffusion coefficients of open heavy flavor mesons in the hadronic medium 
have been studied via the effective 
Lagrangian model and other 
models~\cite{Das:2011vba,Abreu:2011ic,Abreu:2012et,Tolos:2013kva}. 
The charmonium diffusion coefficients, 
in particular for the excited states, have not yet been determined. 
Lacking more rigorous calculations, 
we employ here the geometry scale 
approximation to extract the charmonium diffusion 
coefficients in the hadronic medium.  
Due to the uncertainty in the diffusion coefficients, 
different values are here considered to calculate the $J/\psi$ elliptic flows. 
 
The D meson diffusion coefficient $\mathcal{D}_s^{D}$ 
in the hadronic medium has been calculated 
using the hadron resonance gas (HRG) model. It has been found that 
$4\lesssim \mathcal{D}_s^D(2\pi T)\lesssim 10$ in the temperature region of
$0.8T_c<T<T_c$~\cite{He:2012df}. The value of $\mathcal{D}_s^D (2\pi T)$ 
increases when the temperature decreases in the hadron phase. In this work, we take 
$\mathcal{D}_s^D(2\pi T)=8$ in the temperature region of
$0.8T_c\sim T_c$ for simplicity~\cite{He:2012df}. Additionally, the hadronic medium is assumed to 
reach kinetic freeze-out at 
$T_{\rm kin}=0.8T_c$. 
The charmonium diffusion coefficients are obtained via the geometry scale, 
$\mathcal{D}_s^{\psi} = \mathcal{D}_s^D \times
{{\langle r_{D}\rangle^2\over  \langle r_{\psi}\rangle^2}}$. 
The charge radius of the D meson is approximated to be 
$\langle r_{D}\rangle =0.41$ fm~\cite{Zhao:2020jqu}. The radii of the charmonium ground state and 
excited states ($\chi_c, \psi^\prime$) are 0.5 and 0.72 fm~\cite{Satz:2005hx}, 
respectively. 
Based on this approximation, the charmonium excited states with relatively small spatial 
diffusion coefficients can develop large collective flows in 
the hadronic medium, which will increase the elliptic flows of the prompt and inclusive $J/\psi$ 
at high $p_T$ through the decay process 
$\chi_c(\psi^\prime)\rightarrow J/\psi X$. 

Hot medium evolutions are simulated using the 
$(2+1)$-dimensional ideal 
hydrodynamic equations, and 
the longitudinal expansion is approximated with the Bjorken expansion: 
\begin{align}
\partial_\mu T^{\mu\nu}=0,
\end{align}
where $T^{\mu\nu}= (e+p)u^\mu u^\nu - g^{\mu\nu} p$ is the energy-momentum tensor, while 
$e$ and $p$ are the energy density and pressure, respectively. Corrections due to viscous stresses 
in the hydrodynamic equations have been neglected here~\cite{Shen:2014vra}. 
In the deconfined phase, 
the QGP is treated as an ideal gas of massless $u,d$ quarks and gluons, and 
$s$ quarks with 150 MeV~\cite{Sollfrank:1996hd}. The hadronic medium is treated as an ideal 
gas consisting of hadrons and resonances with mass up to 2 GeV~\cite{Zhu:2004nw,HG:eos}. 
The deconfined and confined phases are combined with 
a first-order phase transition at the critical temperature $T_c=165$ MeV. 
At $\sqrt{s_{NN}}=5.02$ TeV, 
the initial maximum temperatures of the QGP are fixed to be $450$ 
MeV in the forward rapidity $2.5<y<4$ in Pb-Pb collisions~\cite{Chen:2018kfo}.

\section{Elliptic flows of charmonium}
Charmonium production is dominated by the regeneration mechanism 
at low $p_T$, while it is 
dominated by the primordial production at high $p_T$. 
The $J/\psi$ elliptic flows at low $p_T$ are mainly 
due to the kinetic thermalization of charm quarks 
and the regeneration process in the QGP. 
The ALICE collaboration presented a large elliptic flow of inclusive $J/\psi$ 
even at high $p_T$; which is higher than that of the 
theoretical calculations~\cite{Acharya:2017tgv}. The regeneration contribution is negligible at 
high $p_T$. One of the 
factors neglected in the theoretical studies is the charmonium diffusion in 
the hadronic medium. In this work, the hadron phase effects on 
$J/\psi$ collective flows are investigated using the Langevin equation. 
 
\begin{figure}[!hbt]
\centering
\includegraphics[width=0.45\textwidth]{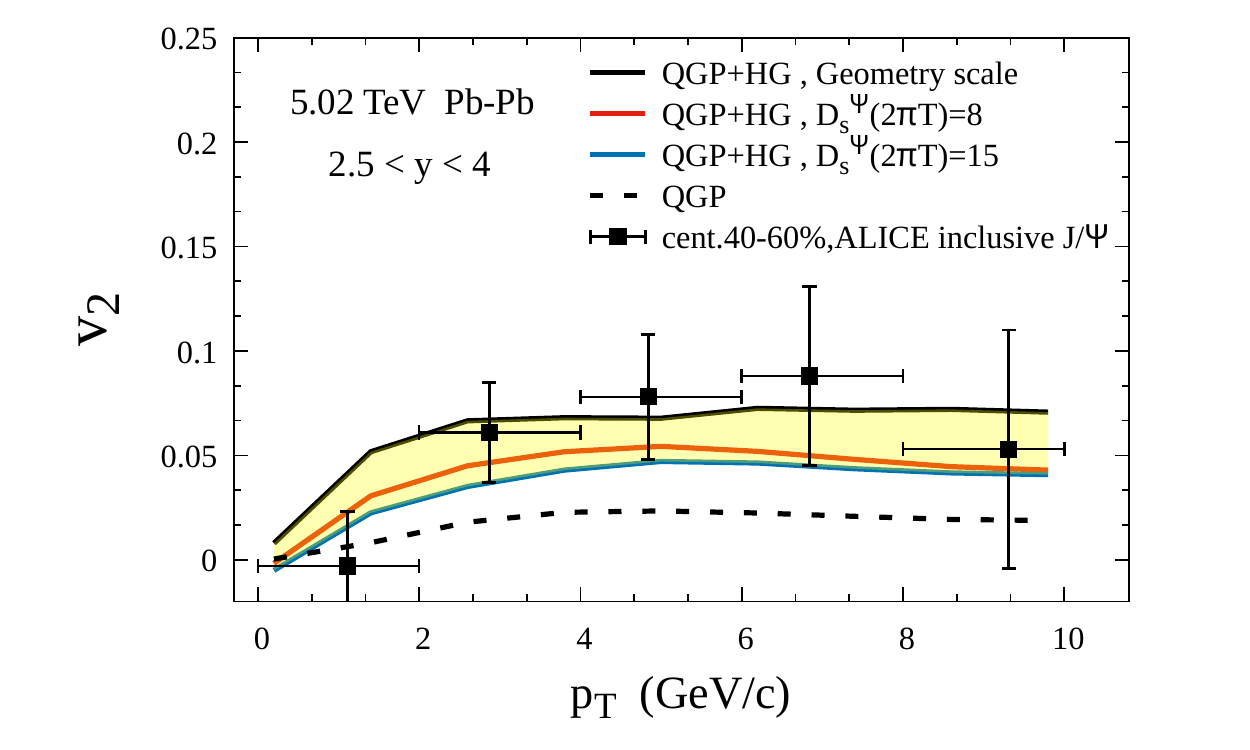}
\caption{(Color online) 
Elliptic flows of the primordial $J/\psi$ as a function of the transverse momentum 
in cent.40-60\% in $\sqrt{s_{NN}}=5.02$ TeV Pb-Pb collisions. The dashed line represents 
$v_2$ of the primordial $J/\psi$ with QGP dissociations. The solid lines include both the 
QGP dissociations and diffusions in the hadronic medium. The diffusion coefficients determined 
by the geometry scale, $\mathcal{D}_s^{\psi}(2\pi T)=8$, and $\mathcal{D}_s^{\psi}(2\pi T)=15$, 
are plotted with black, red, and blue solid lines,  
respectively. 
The decay contributions 
($\chi_c \rightarrow J/\psi X, \psi^\prime\rightarrow J/\psi X$) are included in all lines. 
The regeneration 
and B-hadron decay parts are not included in the theoretical lines. 
The experimental data represents the inclusive $J/\psi$ from the ALICE collaboration~\cite{Acharya:2017tgv}. 
}
\hspace{-0.1mm}
\label{fig-v2-Jpsi}
\end{figure}

The obtained charmonium elliptic flows are plotted in Fig.\ref{fig-v2-Jpsi}. 
The dashed line includes contributions from the cold nuclear matter effects and QGP 
dissociations. The solid lines include contributions from cold nuclear matter effects, 
QGP dissociations, and diffusions in the 
hadronic medium. 
As we focus on the $J/\psi$ elliptic flows at high $p_T$, the regeneration contribution, which 
is mainly located in the $p_T\lesssim 4$ GeV/c region, is deliberately neglected in these 
calculations. 
Based on the geometry scale approximation, 
different charmonium states ($J/\psi, \chi_c, \psi^\prime$) take different values of 
the diffusion coefficients in the hadronic medium. The charmonium excited states can therefore 
obtain larger collective flows from the expanding hadronic medium due to a smaller 
diffusion coefficient. This will increase the $v_2$ of the inclusive $J/\psi$ through the $\chi_c(\psi^\prime)\rightarrow J/\psi X$ decay 
process after the evolutions in the QGP and 
hadron phase, as shown by the black solid line 
in Fig.\ref{fig-v2-Jpsi}. Due to the uncertainty in the charmonium diffusion coefficients, 
different values are also considered (refer to the red and blue lines in Fig.\ref{fig-v2-Jpsi}).  
In the case of the red line, all charmonium states take the same value, i.e.,  
$\mathcal{D}_s^{\psi}(2\pi T)=8$. The hadronic medium effects become smaller. 
In the case of the blue 
line, all charmonium states take a different value, namely $\mathcal{D}^{\psi}(2\pi T)=15$. Both 
these situations yield smaller elliptic flows of the inclusive $J/\psi$, as the 
diffusion coefficients in these two scenarios are larger than 
the value obtained from the geometry scale approximation. 

In the 40-60\% centrality, 
some of the charmonium excited states can survive the QGP dissociations 
and carry elliptic flows from the hadronic medium if their spatial 
diffusion coefficients $\mathcal{D}_s^{\psi}$ are small. 
Considering that 
hadron elastic scatterings can increase the $v_2$ of B mesons 
by $\sim 2\%$~\cite{Cao:2015hia,He:2014cla}, 
a larger effect in the case of charmonium excited states is expected since their masses are 
smaller and the geometry size is larger. 
As Fig.\ref{fig-v2-Jpsi} shows, 
the hadronic effects result in an increase of $v_2^{J/\psi}$ from $\sim 2\%$ to 
$\sim 6\%$ at $p_T\sim 6$ GeV/c when the charmonium diffusion coefficients are extracted using 
the geometry scale approximation. 
It should be noted that at low $p_T$, after including the regeneration contribution, 
the inclusive $J/\psi$ $v_2$ is still dominated 
by the regeneration process. 

\begin{figure}[!hbt]
\centering
\includegraphics[width=0.45\textwidth]{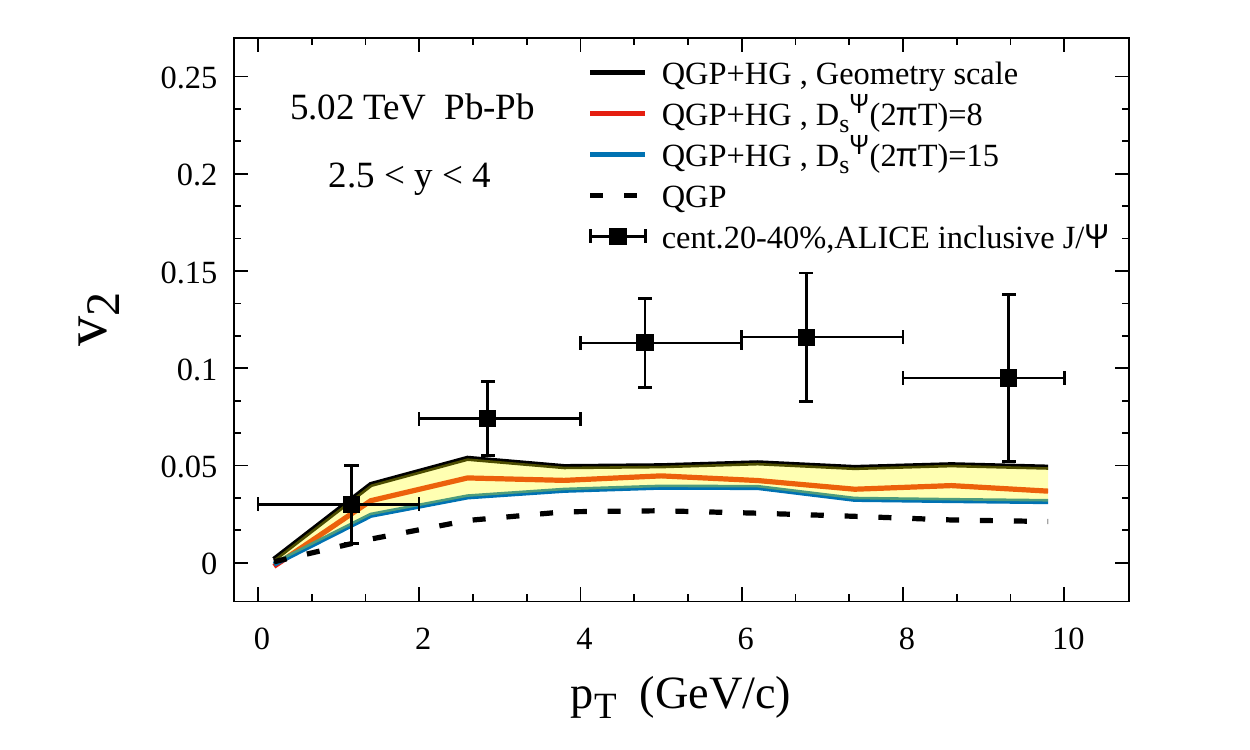}
\caption{ (Color online)  
Elliptic flows of the primordial $J/\psi$ as a function of the transverse momentum 
in cent.20-40\% in $\sqrt{s_{NN}}=5.02$ TeV Pb-Pb collisions. 
The dashed line represents 
$v_2$ of the primordial $J/\psi$ with QGP dissociations. Solid lines include both 
QGP and hadronic medium effects. The other parameters are the same as in Fig.\ref{fig-v2-Jpsi}. 
The experimental data represents the inclusive $J/\psi$ obtained from the ALICE collaboration~\cite{Acharya:2017tgv}. 
}
\hspace{-0.1mm}
\label{fig-v2-Jpsi-b84}
\end{figure}

\begin{figure}[!hbt]
\centering
\includegraphics[width=0.45\textwidth]{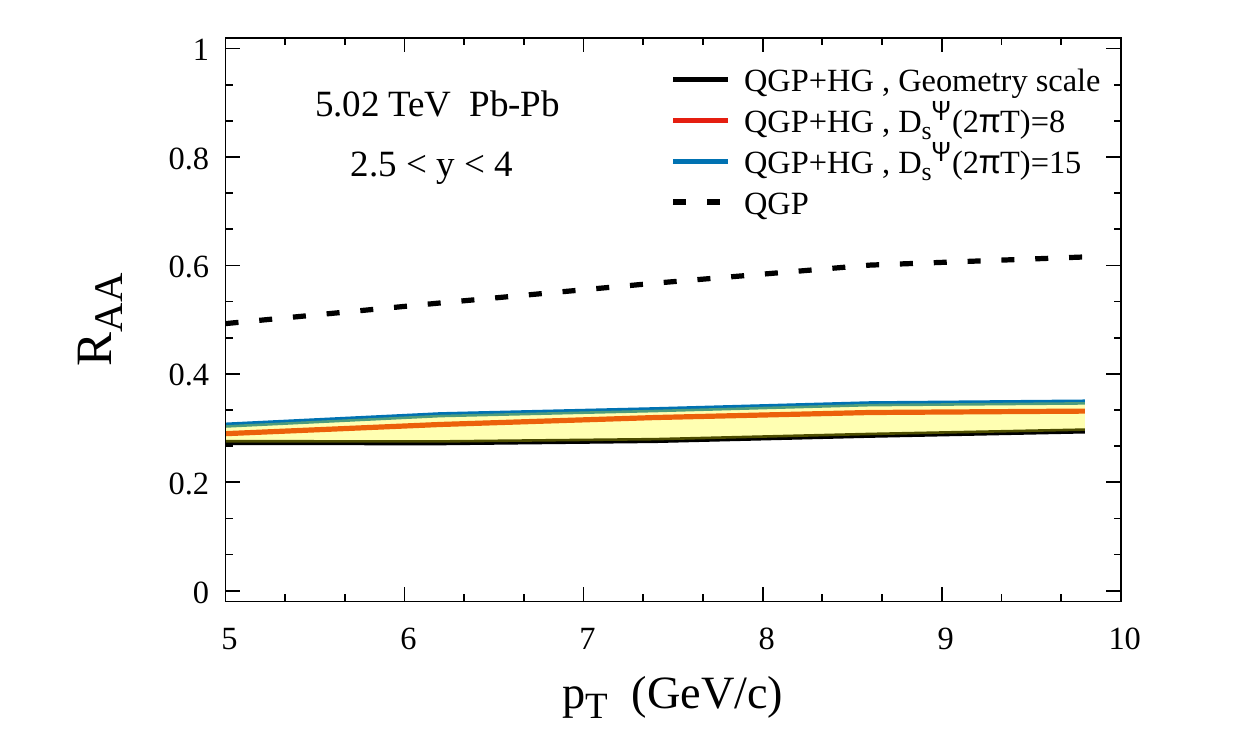}
\caption{ (Color online) $J/\psi$ nuclear modification factor $R_{AA}$ 
as a function of $p_T$ in cent.40-60\%. The other parameters are the same as in Fig.\ref{fig-v2-Jpsi}. 
The dashed line is obtained considering QGP effects only, while the black, red, and blue solid lines 
correspond to the situations where the charmonium diffusion coefficients are determined by 
the geometry scale, 
$\mathcal{D}_s^{\psi}(2\pi T)=8$, and $\mathcal{D}_s^{\psi}(2\pi T)=15$, respectively. 
}
\hspace{-0.1mm}
\label{fig-RAA}
\end{figure}

Fig.\ref{fig-v2-Jpsi-b84} shows the calculations in cent.20-40\%. The hadronic effects result in an 
increase in the elliptic flow of the primordial $J/\psi$ from $\sim 2.5\%$ to $\sim 5\%$. 
The effect is slightly smaller than that observed in cent.40-60\%. This is because 
a larger proportion of the charmonium excited states is dissociated in the QGP, and
their contribution to the final $v_2^{J/\psi}$ is suppressed. In more central collisions 
such as cent.5-20\%, the hadronic effects become smaller than those illustrated in Figs.\ref{fig-v2-Jpsi}-\ref{fig-v2-Jpsi-b84}. 
The addition of the hadron phase 
leads to an enhancement of $v_2$ of the primordial $J/\psi$; furthermore, this addition reduces 
the nuclear modification factor $R_{AA}$ of $J/\psi$ at $p_T\gtrsim5$ GeV/c 
where the primordial production dominates, 
as shown in Fig.\ref{fig-RAA}. 

\begin{figure}[!hbt]
\centering
\includegraphics[width=0.42\textwidth]{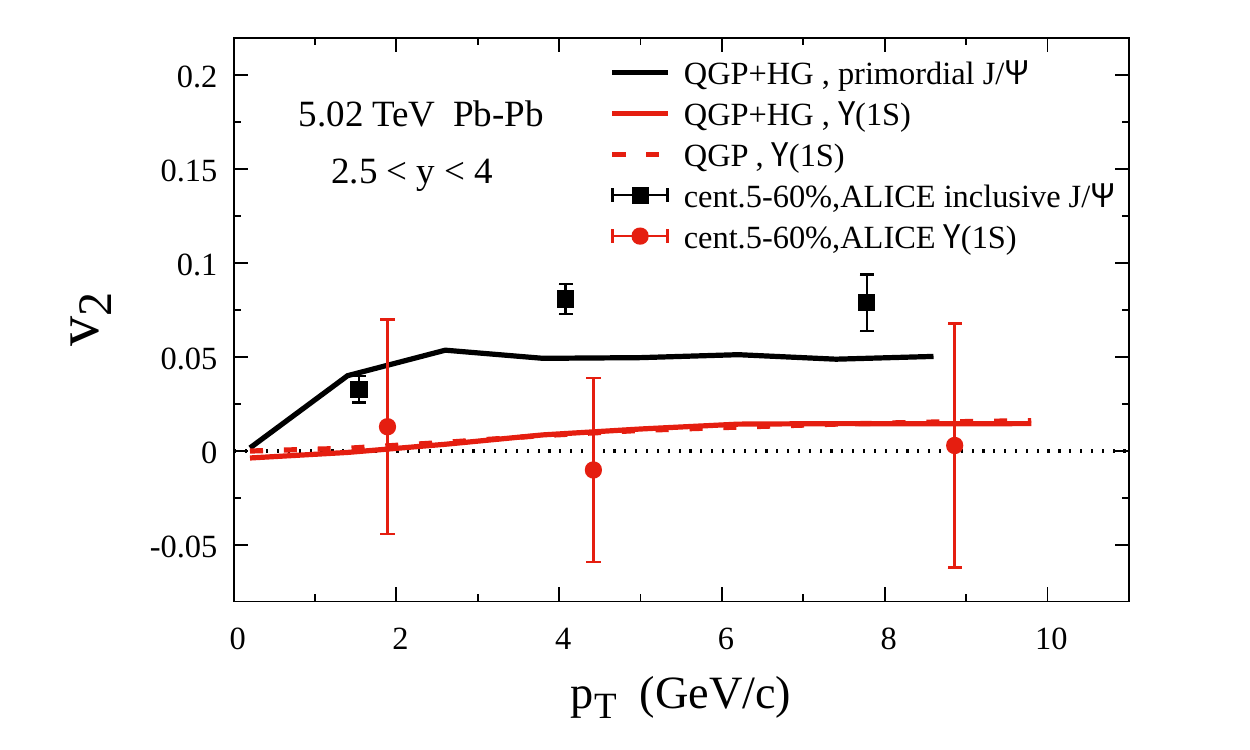}
\caption{ (Color online) 
Elliptic flows of the inclusive $J/\psi$ and $\Upsilon$(1S) as a function of $p_T$ in the 
forward rapidity $2.5<y<4$ in cent.5-60\% in $\sqrt{s_{NN}}=5.02$ TeV Pb-Pb collisions. 
The theoretical calculations of $J/\psi$ are the same as those presented in 
Fig.\ref{fig-v2-Jpsi-b84}. The diffusion coefficients of $J/\psi$ and $\Upsilon(1S)$ 
are obtained using the geometry scale. 
Regeneration is absent here. 
The experimental data are obtained from the ALICE 
collaboration~\cite{Acharya:2019hlv}.    
}
\hspace{-0.1mm}
\label{fig-v2-upsilon}
\end{figure}

Fig.\ref{fig-v2-upsilon} shows 
the hadronic effects on $\Upsilon$(1S) (red solid line). 
As bottomonia are tightly bound states, the QGP dissociations contribute only 
$\sim 1\%$ to the $v_2$ of $\Upsilon$(1S) through
path length difference effect~\cite{Bhaduri:2018iwr}. 
The difference between the red solid line and the red dashed line in Fig.\ref{fig-v2-upsilon} 
represents the effect of the hadronic 
medium on the $v_2$ of  
$\Upsilon$(1S); this effect is negligible due to the upsilon large mass and 
small geometry size. 
In addition, the substantial difference between the elliptic flows of $J/\psi$ and $\Upsilon$(1S) in Fig.\ref{fig-v2-upsilon} indicates that the former 
come from the final state interactions rather than from the initial cold nuclear matter 
effects~\cite{Zhang:2019dth}. 
This is beneficial to the understanding of the origin of the charmonium collective behaviors in the hot 
medium.  
  
\section{Summary }

The transport and Langevin equations were employed to study the elliptic flows of 
primordial charmonium in the QGP and hadron phase. 
Charmonium diffusions in the hadronic medium were simulated using the Langevin equation. 
When the spatial diffusion coefficients in the Langevin equation are parametrized via the 
geometry scale approximation, 
the charmonium $v_2$ can be enhanced by $\sim 5\%$ at high $p_T$ after considering 
the hadronic medium effects. This contribution would be suppressed if a larger value of the  
diffusion coefficient was considered. The $J/\psi$ nuclear modification factor 
at high $p_T$ is also suppressed by the diffusion process. 
The bottomonium $v_2$ is not substantially affected by these 
hadronic medium effects due to the large bottomonium mass. The difference between the   
charmonium and bottomonium experimental data indicates that the elliptic flows of charmonium 
at high $p_T$ come from the final state interactions in the hot medium. 
The study of charmonium diffusions in the hadronic medium 
aids the understanding of the experimental data of quarkonium 
collective flows in ultrarelativistic heavy ion collisions.

\vspace{0.8cm}
\appendix {{\bf Acknowledgements:} BC is grateful to Wenbin Zhao for discussions 
regarding the hydrodynamic model. This work is supported by the NSFC
Grant No. 11705125 and the ``Qing-Gu" project (2019XRG-0066) of Tianjin University.}

\end{document}